\documentclass[prb,twocolumn]{revtex4}%

\newcommand{\be}{\begin{equation}}
\newcommand{\ee}{\end{equation}}
\newcommand{\bea}{\begin{eqnarray*}}
\newcommand{\eea}{\end{eqnarray*}}
\newcommand{\bean}{\begin{eqnarray}}
\newcommand{\eean}{\end{eqnarray}}

\begin{document}

\draft
\title
{\bf  Thermal rectification effects of multiple semiconductor
quantum dot junctions}

\author{ David M.-T. Kuo }

\address{Department of Electrical Engineering and Department of Physics, National Central
University, Chungli, 320 Taiwan}


\date{\today}

\begin{abstract}
Based on the multiple energy level Anderson model, this study
theoretically examines the thermoelectric effects of semiconductor
quantum dots (QDs) in the nonlinear response regime. The charge
and heat currents in the sequential tunneling process are
calculated by using the Keldysh Green's function technique.
Results show that the thermal rectification effect can be observed
in a multiple QD junction system, whereas the tunneling rate, size
fluctuation, and location distribution of QD significantly
influence the rectification efficiency.
\end{abstract}

\maketitle
\section{Introduction}

Due to global warming and environment issues, it is important to
developing the new energies to reduce the total amount of $CO_2$
in the earth's atmosphere. Electrical power generated by the
mechanical force of wind and ocean tides is one of several types
of green energy. Apart from that, electrical power generated from
solar energy transformation also creates clean energy. Researchers
have used many technologies to develop  materials with high values
of figure of merit (ZT) for potential applications of solid-state
thermal devices. These materials can be utilized to generate
electrical power through the temperature bias resulting from solar
heating or other method heating.[1-9]

Researcher have proposed several methods for enhancing ZT.[2] One
of them is to reduce system dimensions.[3] A zero dimension QD
system was predicted to be more pronounced for the enhancement of
thermoelectric efficiency in the dimension reduction.[9]
Nevertheless, previous studies focus on the linear response regime
of $\Delta T/T_0 \ll 1$, where $\Delta T$ and $T_0$ are the
temperature bias and equilibrium temperature, respectively, of two
side electrodes.[10] The properties of thermal devices in the
nonlinear response regime $\Delta T/T_0 \sim 1$ have important
applications such as thermal rectifiers and thermal transistors. A
thermal rectifier is crucial for the heat current storage. Records
of thermal rectification date back as early as 1935, when Starr
discovered that copper oxide/copper junctions can exhibit a
thermal diode behavior.[11] The thermal rectification effects have
been theoretically predicted to occur in one dimensional phonon
junction systems.[12-15]  Because phonons are heat flow carriers
in refs [12-15], their applications are restricted to the heat
manipulation and heat energy storage. Using charge carriers
(electrons or holes), we can manipulate heat currents and charge
currents.  The Peltier effect with the application of coolers
makes it possible to manipulate the heat current using applied
voltage. Its reverse process is so called the Seebeck effect used
to create electrical power.

Scheibner and coworkers have, recently, employed a single metallic
QD junction to measure electrochemical potential in the linear
response regime.[16] Such a small electrochemical potential
yielded by temperature bias is not sufficient to demonstrate the
thermal rectification effect.[17] Besides, the operation
temperature of a metallic QD is very low because its charging
energies are much smaller than the thermal energy
($k_BT_0=25~meV$). Semiconductor QD junctions with large interdot
Coulomb interactions and energy-level separation within each QD
allow semiconductor QD thermal rectifiers to operate at much
higher temperatures. Previous studies of refs[16,17] only consider
a single QD junction, which is useful for the case of low QD
density. To scale up QD thermal devices, it is necessary to
consider the case of high QD density.

This study demonstrates that a system of multiple semiconductor
QDs embedded in an amorphous insulator with low heat conductivity
connected to the metallic electrodes exhibits the thermal
rectification effect in the nonlinear response regime. This study
also clarify the thermal power in the linear and nonlinear
regimes. Although the mechanism of thermal rectification for QD
junctions is similar to that of charge current, the heat current
is yielded by temperature bias and electrochemical potential. The
very nonlinear relation between the applied temperature bias and
the electrochemical potential leads to that it is not
straightforward to reveal the relation between the heat current
and the applied temperature bias.

\section{Formalism}

The following multi-level Anderson model [18] describes the
Hamiltonian of the metal/quantum dots/metal double barrier
junction in Fig. 1 ;

\bea H& =&\sum_{k,\sigma,\beta} \epsilon_k
a^{\dagger}_{k,\sigma,\beta}a_{k,\sigma,\beta}+\sum_{\ell,\sigma}
E_{\ell} d^{\dagger}_{\ell,\sigma} d_{\ell,\sigma}
\nonumber \\
 &+& \sum_{\ell,\sigma}
U_{\ell} d^{\dagger}_{\ell,\sigma} d_{\ell,\sigma}
d^{\dagger}_{\ell,-\sigma} d_{\ell,-\sigma} +\frac{1}{2}\sum_{\ell
\neq j;\sigma,\sigma'}
U_{\ell,j} d^{\dagger}_{\ell,\sigma} d_{\ell,\sigma} d^{\dagger}_{j,\sigma'} d_{j,\sigma'} \\
\nonumber &+&\sum_{k,\sigma,\beta,\ell}
V_{k,\beta,\ell}a^{\dagger}_{k,\sigma,\beta}d_{\ell,\sigma}+
\sum_{k,\sigma,\beta,\ell}
V^{*}_{k,\beta,\ell}d^{\dagger}_{\ell,\sigma}a_{k,\sigma,\beta}
\eea where $a^{\dagger}_{k,\sigma,\beta}$ ($a_{k,\sigma,\beta}$)
creates (destroys) an electron of momentum $k$ and spin $\sigma$
with energy $\epsilon_k$ in the $\beta$ metallic electrode. The
term $d^{\dagger}_{\ell,\sigma}$ ($d_{\ell,\sigma}$) creates
(destroys) an electron with the ground-state energy $E_{\ell}$ of
$\ell$th QD , while $U_{\ell}$ and $U_{\ell,j}$ describe,
respectively, the intradot Coulomb interactions and the interdot
Coulomb interactions . The term $V_{k,\beta,\ell}$ describes the
coupling between the band states of electrodes and the QDs. The
Hamiltonian of Eq. (1) assumes that energy level separations
between the ground state and the first excited state within each
QD is much larger than intradot Coulomb interactions $U_{\ell}$
and thermal temperature $k_BT$. Therefore, there is only one
energy level for each QD. This study ignores the interdot hopping
terms due to the high potential barrier separating QDs.

Using the Keldysh-Green's function technique,[19] the charge and
heat currents leaving electrodes can be expressed as \bea
J_e&=&\frac{-2e}{h}\sum_{\ell} \int d\epsilon
\gamma_{\ell}(\epsilon) ImG^r_{\ell,\sigma}(\epsilon)
f_{LR}(\epsilon), \eea

\bea Q &=&\frac{-2}{h}\sum_{\ell} \int d\epsilon
\gamma_{\ell}(\epsilon)
ImG^r_{\ell,\sigma}(\epsilon)(\epsilon-E_F-e\Delta V)
f_{LR}(\epsilon) , \eea where
$f_{LR}(\epsilon)=f_{L}(\epsilon)-f_{R}(\epsilon)$, the
transmission factor is
$\gamma_{\ell}(\epsilon)=\frac{\Gamma_{\ell,L}(\epsilon)
\Gamma_{\ell,R}(\epsilon)}
{\Gamma_{\ell,L}(\epsilon)+\Gamma_{\ell,R}(\epsilon)}$,
$f_{L(R)}(\epsilon)=1/(exp^{(\epsilon-\mu_{L(R)})/(k_BT_{L(R)})}+1)$
is the Fermi distribution functions for the left (right) electrode
. The chemical potential difference between these two electrodes
is related to the bias difference $\mu_{L}-\mu_{R}=e \Delta V$.
The temperature difference is $T_L-T_R=\Delta T$, and $E_F$ is the
Fermi energy of electrodes. $\Gamma_{\ell,L}(\epsilon)$ and
$\Gamma_{\ell,R}(\epsilon)$ [$\Gamma_{\ell,\beta}=2\pi\sum_{{\bf
k}} |V_{\ell,\beta,{\bf k}}|^2 \delta(\epsilon-\epsilon_{{\bf
k}})]$ denote the tunneling rates from the QDs to the left and
right electrodes, respectively. Notations $e$ and $h$ denote the
electron charge and Plank's constant, respectively. For
simplicity, we assumed that these tunneling rates are energy- and
bias-independent. Equations (2) and (3) have been employed to
reveal the thermal properties of a single-level QD in the Kondo
regime.[20,21] Here, this analysis focuses on the heat current in
the Coulomb blockade regime. Previous studies [22,23] provide an
expression of the retarded Green function

\begin{eqnarray}
G^{r}_{\ell,\sigma}(\epsilon)&=&(1-N_{\ell.-\sigma})\sum^{3^{n-1}}_{m=1}
\frac{p_m}{\epsilon-E_{\ell}-\Pi_m+i\Gamma_{\ell}} \nonumber \\
 &+& N_{\ell.-\sigma}\sum^{3^{n-1}}_{m=1}
\frac{p_m}{\epsilon-E_{\ell}-U_{\ell}-\Pi_m+i\Gamma_{\ell}},
\end{eqnarray}
where $n$ denotes the number of QDs. The term $\Pi_m$ denotes the
sum of Coulomb interactions seen by a particle in dot $\ell$ due
to other particles in the dot $j (j\ne \ell)$, which can be
occupied by zero, one or two particles. The probability factor
$p_m$ denotes the probability of such configurations.  For a
three-QD system ($\ell \neq j \neq j'$), there are nine ($3\times
3$) configurations, and the probability factors become $p_1=a_j
a_{j'}$, $p_2=b_j a_{j'}$, $p_3=a_{j} b_{j'}$, $p_4=c_j a_{j'}$,
$p_5=c_{j'}a_{j}$, $p_6=b_j b_{j'}$, $p_7=c_j b_{j'}$, $p_8=c_{j'}
b_{j}$, and $p_9=c_{j} c_{j'}$, where
$a_j=1-(N_{j,\sigma}+N_{j,-\sigma})+c_j$,
$b_j=(N_{j,\sigma}+N_{j,-\sigma})-2c_j$, and $c_j=\langle
n_{j,-\sigma}n_{j,\sigma}\rangle$ is the intradot two particle
correlation function. $N_{j,\sigma}$ is one particle occupation
number. The interdot Coulomb interaction factors are $\Pi_1=0$,
$\Pi_2=U_{\ell,j}$, $\Pi_3=U_{\ell,j'}$, $\Pi_4=2U_{\ell,j}$,
$\Pi_5=2U_{\ell,j'}$, $\Pi_6=U_{\ell,j}+U_{\ell,j'}$,
$\Pi_7=2U_{\ell,j}+U_{\ell,j'}$, $\Pi_8=2U_{\ell,j'}+U_{\ell,j}$,
and $\Pi_9=2U_{\ell,j}+2U_{\ell,j'}$. The sum of probability
factors $p_m$ for all configurations equals to 1, indicating the
fact that $G^{r}_{\ell,\sigma}(\epsilon)$ satisfies the sum rule.
The imaginary part of each resonant channel is
$\Gamma_{\ell}=(\Gamma_{\ell,L}+\Gamma_{\ell,R}))/2$ resulting
from the coupling between the QDs and the electrodes, where the
real part of self energy is ignored. A self energy that ignores
the effects of electron Coulomb interactions is adequate in the
Coulomb blockade regime, but it does not capture the Kondo effect.

According to the expression of the retarded Green's function in
Eq. (4), we need to know the occupation numbers
$N_{\ell,\sigma}(N_{\ell,-\sigma})$ and $N_{\ell,\ell}=c_{\ell}$.
These numebrs can be obtained by solving the following equations
self-consistently:
\begin{small}
\begin{equation}
N_{\ell,\sigma} = -\int \frac{d\epsilon}{\pi}
\frac{\Gamma_{\ell,L} f_{L}(\epsilon)+\Gamma_{\ell,R}
f_{R}(\epsilon)}{\Gamma_{\ell,L}+\Gamma_{\ell,R}}
ImG^r_{\ell,\sigma}(\epsilon)
\end{equation}
\end{small}
\begin{eqnarray}
c_{\ell} = -\int \frac{d\epsilon}{\pi} \frac{\Gamma_{\ell,L}
f_{L}(\epsilon)+\Gamma_{\ell,R}
f_{R}(\epsilon)}{\Gamma_{\ell,L}+\Gamma_{\ell,R}} Im
G^{r}_{\ell,\ell}(\epsilon)
\end{eqnarray}
The values of $N_{\ell,\sigma}$ and $c_{\ell}$ are restricted
between 0 and 1. The expression of two particle retarded Green
function of Eq. (6) is

\[
G^r_{\ell,\ell}(\epsilon)=N_{\ell.-\sigma}\sum^{3^{n-1}}_{m=1}
\frac{p_m}{\epsilon-E_{\ell}-U_{\ell}-\Pi_m+i\Gamma_{\ell}}.
\]

\section{Results and discussion}


Although the thermal rectification effect of a single QD junction
with two levels has been theoretically investigated in ref.[17],
it is difficult to realize the physical condition of asymmetrical
coupling between the QD and the electrodes, where the excited
state states have symmetrical tunneling rates, but the ground
state has an asymmetrical tunneling rate. Meanwhile, ref[17] did
not take into account the crucial intralevel Coulomb interactions.
To determine the direction-dependent heat current, we considered
two sets of parameters (the cases of $T_L
> T_R$ and $T_R >
T_L$): (a) is $\mu_L=E_F+e\Delta V$, $\mu_R=E_F$, $T_L=T_0+\Delta
T$, and $T_R=T_0$, and (b) is $\mu_L=E_F$, $\mu_R=E_F-e\Delta
V$,$T_L=T_0$ and $T_R=T_0+\Delta T$. The average temperature of
two side electrodes is $(T_L+T_R)/2=T_0+\Delta T/2$, which is
increasing steadily with $\Delta T$. Such a physical condition for
average temperature is more readily realized than that of
reference [17] from experimental point of view.[24] A functional
thermal rectifier requires a good thermal conductance for the case
of $T_L
> T_R$, but a thermal insulator for the case of $T_R
> T_L$. Based on Eqs. (2) and (3), a QD thermal rectifier
requires not only highly asymmetrical coupling strengthes between
the QDs and the electrodes, but also strong electron Coulomb
interactions between the dots.  To investigate the thermal
rectification behavior, we numerically solved Eqs. (2) and (3) for
three QD junctions for using various system parameters.

The energy level of dot $\ell$ is
$E_{\ell}=E_F+\alpha_{\ell}\Delta E$, where $-1 \le \alpha_{\ell}
\le 1$. The term $\Delta E=200\Gamma$ represents the energy level
fluctuation of QDs. For simplicity, the intradot Coulomb
interactions $U_{\ell}=600\Gamma$ are fixed, although the QD size
fluctuation leads to the intradot Coulomb interaction variation.
According to Eq. (3), the direction-dependent heat currents are
attributed to $\Delta V$ and
$N_{\ell,\sigma}=N_{\ell,-\sigma}=N_{\ell}$. The Seebeck effect
for the open circuit ($J_e=0$) determines the electrochemical
potential $e\Delta V$. This electrochemical potential forms in
response to the current generated by the temperature gradient.
This electrochemical potential is the Seebeck voltage (Seebeck
effect). The thermal power is $S=\Delta V/\Delta T$. Once $\Delta
V$ is solved, we then use Eq. (3) to compute the heat current.

Figure. 2 shows the direction-dependent heat current (Q) and
rectification efficiency ($\eta_Q$) as functions of $\Delta T$ for
various values of $\Gamma_{AR}$, while keeping
$\Gamma_{B(C),L}=\Gamma_{B(C),R}=\Gamma$. We adopted the interdot
Coulomb interactions $U_{\ell,j}=300\Gamma$ and $k_BT_0=25\Gamma$.
The energy levels of dots A, B, and C were chosen to be
$E_A=E_F-\Delta E/5$, $E_B=E_F+2\Delta E/5$ and $E_C=E_F+\Delta
E/5$. The heat currents have positive and negative signs
corresponding for the cases of $T_L > T_R $ and $T_R
> T_L$, respectively. The former is the heat current from the left electrode to
the right electrode $Q_{LR}$, while the later is the heat current
from the right electrode to the left electrode $Q_{RL}$. The
rectification efficiency is $\eta_Q=(Q_{LR}(\Delta
T)-|Q_{RL}(\Delta T)|)/Q_{LR}(\Delta T)$. The rectification
effects of heat currents in Fig. 2(a) are clearly distinguishable
from Fig. 2(b). When dot A is fully blocked ($\Gamma_{AR}=0$), the
QD junction exhibits a much higher $\eta_Q$ and the heat currents
decrease dramatically. This serious suppression implies that the
heat currents through dots B and C are small because their
resonant energy levels are far away from the Fermi energy level
due to a large $U_{AC}$ and $U_{AB}$. The results in Fig. 2
indicate that precisely manipulating the coupling strengthes
between the dots and the electrodes is crucial to the optimization
of thermal rectifiers.

Figure. 2 illustrates the homogenous interdot Coulomb
interactions. However it is difficult to achieve homogenous
interdot Coulomb interactions in the multiple QD layout. The
interdot Coulomb interaction fluctuation mainly arises from the
size variation and the inhomogenous position distribution.
Therefore, we examined the variation effect of $U_{AC}$ on heat
currents. Figure. 3 shows the heat current and rectification
efficiency as functions of $\Delta T$ for various values of
$U_{AC}$ at $\Gamma_{AR}=0$ and $\Gamma_{AL}=2\Gamma$. The other
parameters are the same as those in Fig. 2. In this case, large
interdot Coulomb interactions suppress the heat currents. This
behavior is similar to the Coulomb blockade effect on the charge
current.$^{22,23}$ Nevertheless, a strong $U_{AC}$ leads to higher
rectification efficiency. Therefore, a higher efficient thermal
rectification effect appears in the high density QD system. In the
such a case of strong intradot and interdot Coulomb interactions,
the charge current exhibits clear plateaus with respect to applied
voltage. However, the structure of heat currents in Fig. 3 does
not exhibit these plateaus with respect to temperature bias. This
is because temperature bias broadens the electron
distribution.[22]

In Figures (2) and (3) we illustrate the case of $E_C < E_B$.
Figure. 4 shows that $E_C$ influences the heat current and
$\eta_Q$. The other parameters are the same as those in Fig. 3.
When $E_C$ changes from $E_C=E_F+\Delta E/5$ to $E_C=E_F+4\Delta
E/5$, the heat current for $E_C=E_F+2\Delta E/5$ is less than that
for $E_C=E_F+3\Delta E/5$. This indicates that the separation
between the energy levels of QDs and the Fermi energy level is not
a unique factor to influence the heat current magnitude. When
$E_C$ increases to $E_C=E_F+4\Delta E/5$ from $E_C=E_F+3\Delta
E/5$, the heat currents and rectification efficiency change
slightly. All of these complicated behaviors are the result of the
very nonlinear relationship between the heat currents and
electrochemical potentials, which are strongly influenced by the
electron Coulomb interactions and the energy levels of each QD.

The following expression of heat current, $Q=Q_{B}+Q_{C}$, further
clarifies the results of Fig. 4:
\begin{small}
\begin{eqnarray}
Q_B/(\gamma_B \pi)&=&(1-N_B)*((1-2N_A)(E_B-E_F)f_{LR}(E_B)\\
\nonumber
&+&2N_A(E_B+U_{AB}-E_F)*f_{LR}(E_B+U_{AB})\\
\nonumber Q_C/(\gamma_C
\pi)&=&(1-N_C)*((1-2N_A)(E_C-E_F)f_{LR}(E_C)\\&+&2N_A(E_C+U_{AC}-E_F)*f_{LR}(E_C+U_{AC}),
\end{eqnarray}
\end{small}

Equations. (7) and (8) are the results of setting $c_{\ell}$ to
zero due to very large intradot Coulomb interactions. Further, a
delta function replaces the Lorentzian function of resonant
channels for $\Gamma \ll k_B T_0$ in Eq. (3). In addition, Eqs.
(7) and (8) only consider the interdot Coulomb interactions
$U_{AB}$ and $U_{AC}$ since there is one particle occupation in
dot A. As for $N_A$, $N_B$, $N_C$ and $e\Delta V$ in Eqs. (7) and
(8), we use full numerical solution (Eqs (2),(5) and (6)) to solve
them. Figure. 5 shows the heat current as functions of $\Delta T$
for $E_C=E_F+3\Delta E/5$. The other parameters are the same as
those in Fig.4. The lot lines given in Eqs. (7) and (8) match the
full solutions very well. $Q_{LR}$ and $Q_{RL}$ are the nonlinear
functions of $\Delta T$. Figure. 5(b) shows the $Q_B$ and $Q_C$
calculated by results of Eqs. (7) and (8) for $Q_{LR}$ and
$Q_{RL}$. Based on Fig. 5(b), $Q_C$ is the primary determinant of
the heat currents of Fig. 5(a). The factors of $N_A$ and $f_{LR}$
influence the behavior of $Q_{LR}$ with respect to $\Delta T$.
Only the factor of $f_{LR}$ influences the behavior of $Q_{RL}$.
This is because of $\Delta T$-independent $N_A=0.45$ in the case
of $Q_{RL}$. So far, we have analyzed the heat current and
$\eta_Q$. From experimental point of view, it is easy to measure
the electrochemical potential yielded by the temperature bias
although it is not straightforwardly related to the heat current
in this system.

Figure. 6 shows the electrochemical potential of different values
of $k_BT_0$. The other parameters are the same as those in Fig. 5.
The dot and dashed lines represent the linear functions of $\Delta
T$, whereas the solid lines are nonlinear with respect to $\Delta
T$. Although the asymmetrical electrochemical potentials (Fig.
(6a) and (6b)) reveal some information about thermal
rectification, it is difficult to determine the rectification
efficiency from the measurement of electrochemical potentials.
This is true for both linear and nonlinear responses due to the
very nonlinear relation between Q and $\Delta V$. Therefore, it is
difficult to judge the thermal rectification effect from the
experimental results done in ref[16]. The large electrochemical
potential created by the temperature bias $(\Delta T)$ is useful
for generating electrical powers. The results of Fig. 6 also imply
that the charge current rectification with respect to temperature
appears in the closed circuit ($J_e \neq 0$). An electrochemical
potential is an essential quantity for charge carriers to deliver
heat current, it is different from the phonon carriers in a phonon
junction system.$^{12-15}$ To further illustrate the strong
nonlinear relationship between electrochemical potential and heat
current, we show the heat current and rectification efficiency in
Fig. 7, where the curves exhibit one to one correspondence with
those in Fig. 6. We note that $\eta_Q$ is enhanced when
temperature $k_BT_0 $ decreases. A comparison of Fig. 6 and Fig. 7
demonstrates the strong nonlinear relationship between Q and
$\Delta V$.

Finally, figure 8 calculates the thermal power as a function of
$k_BT_0$ for different values of $\Delta T$. The other parameters
are the same as those in Fig. 5. As expected, the temperature bias
has a significant influence on thermal powers at low $k_BT_0$, but
not at high temperatures. Note that the thermal power values
increase with increasing temperature bias. Thermal power plays a
significant role in determining the figure of merit.[10] Some
thermal devices such as solid state coolers are in the nonlinear
response regime.$^{2}$ Therefore, the previous studies of linear
response regime underestimate the thermal power values when
$\Delta T/T_0 \ge 1$.[10,25]

\section{Conclusions}

This theoretical study reports the thermal rectification effects
(TRE) of multiple QDs embedded into an amorphous insulator (phonon
glass) with low heat conductivity. Results show that the
asymmetrical tunneling rates and strong interdot Coulomb
interactions significantly influence the TRE. Recent research
indicates that it is possible to precisely manipulate the size and
position of multiple germanium QDs in $SiO_2$ amorphous
insulator.[26] This experiment indicates that it is possible to
realize the system shown in Fig. 1 to examine the thermal
rectification at high temperatures. To avoid the phonon heat
current, which can seriously suppress the heat rectification
effect arising from electrons, scan-tunneling- microscopes (STMs)
can replace one of the electrodes since the vacuum layer between
the STM and amorphous insulator can blockade the phonon heat
current.[27]

{\bf Acknowledgments}

This work was supported by the National Science Council of the
Republic of China under Contract Nos. NSC 97-2112-M-008-017-MY2
and NSC 98-2112-M-001-022-MY3.


\mbox{}

\newpage

{\bf Figure Captions}

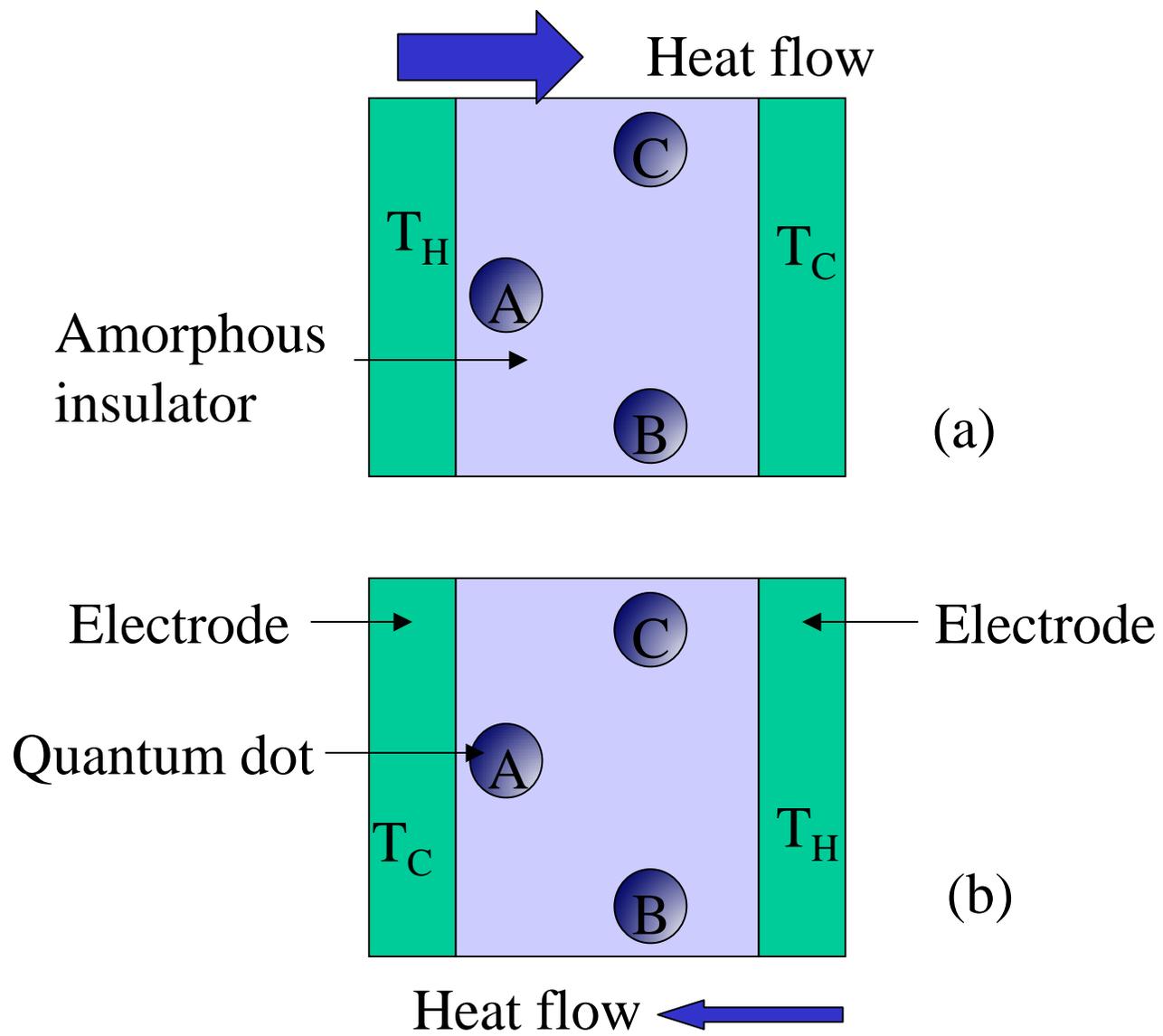
Fig. 1. Schematic representation of multiple semiconductor quantum
dots embedded into an amorphous insulator with low heat
conductivity sandwiched between metallic electrodes.

Fig. 2. Heat current and rectification efficiency as functions of
$\Delta T$ for the tunneling rate variation of dot A. The heat
currents are in the units of $Q_0=\Gamma^2/(2h)$. Such units are
used through out this article.

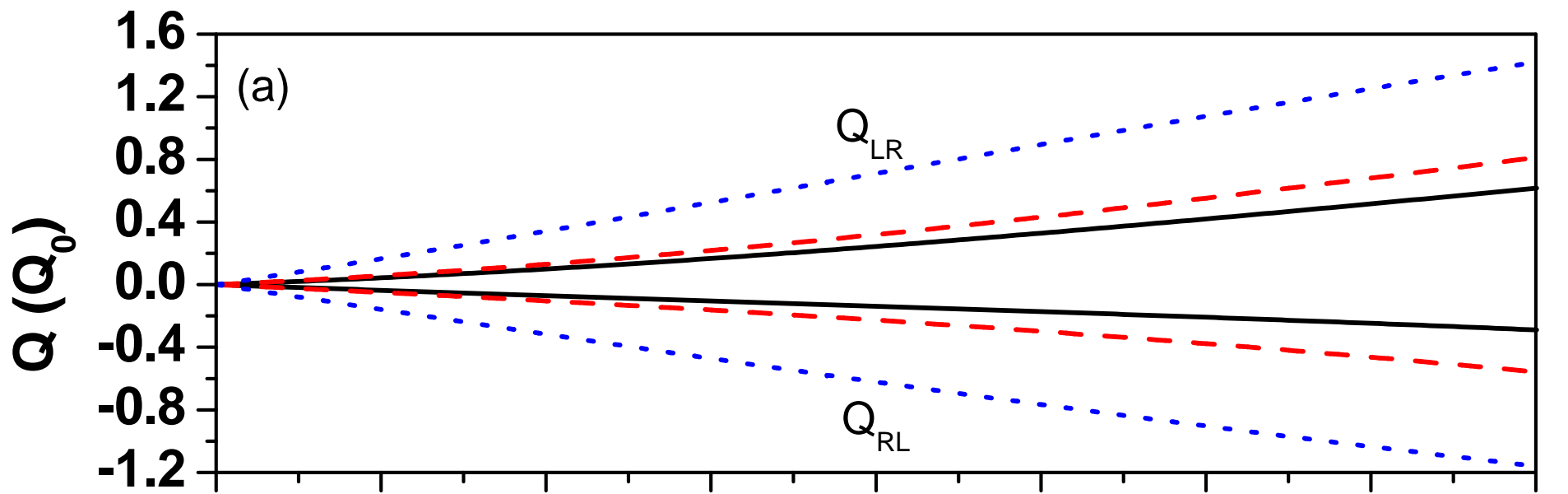
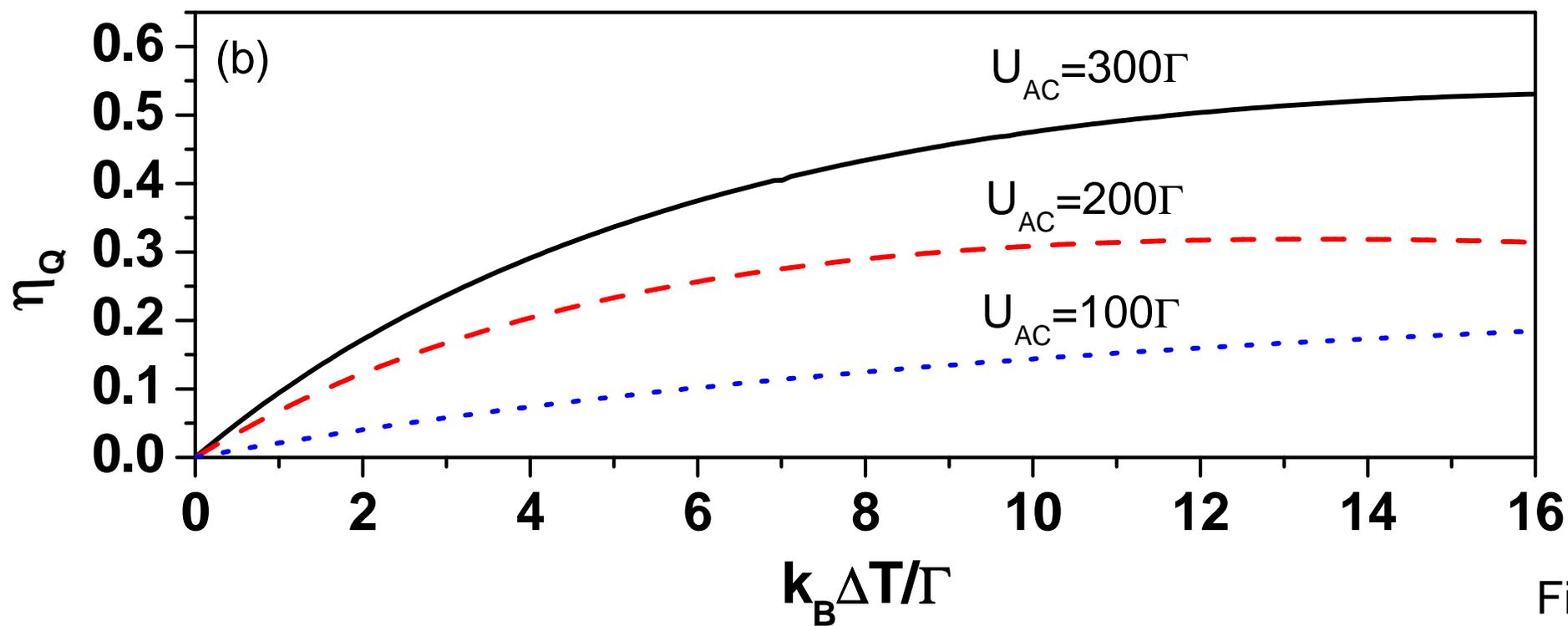
Fig. 3. Heat current and rectification efficiency as functions of
$\Delta T$ for different interdot Coulomb interactions of $U_{AC}$
at  $\Gamma_{AR}=0$ and $\Gamma_{AL}=2\Gamma$. The other
parameters are the same as those in Fig. 2.

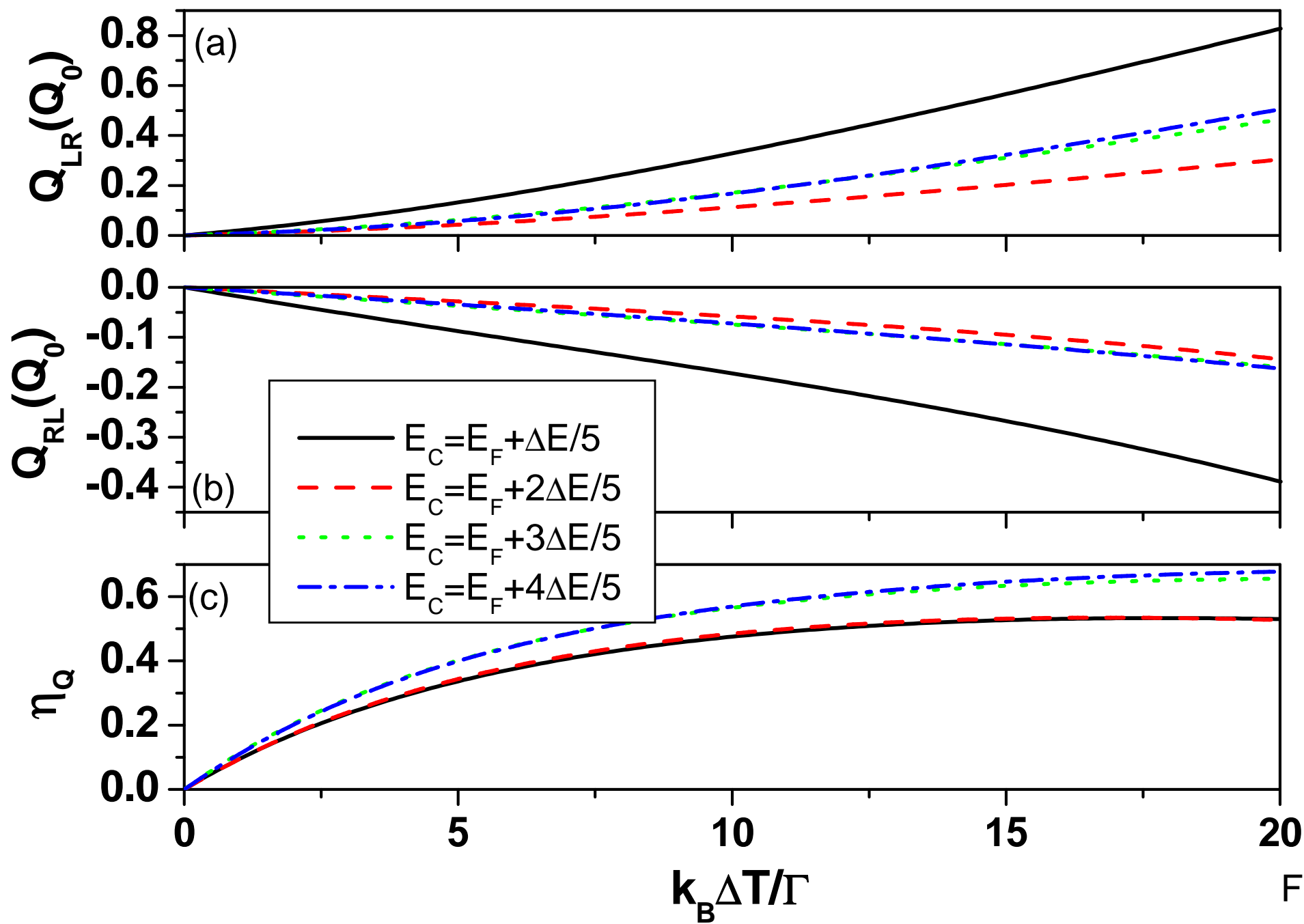
Fig. 4. Heat current and rectification efficiency as functions of
$\Delta T$ for different energy levels of $E_C$ and
$U_{AC}=300\Gamma$. The other parameters are the same as those in
Fig. 3.

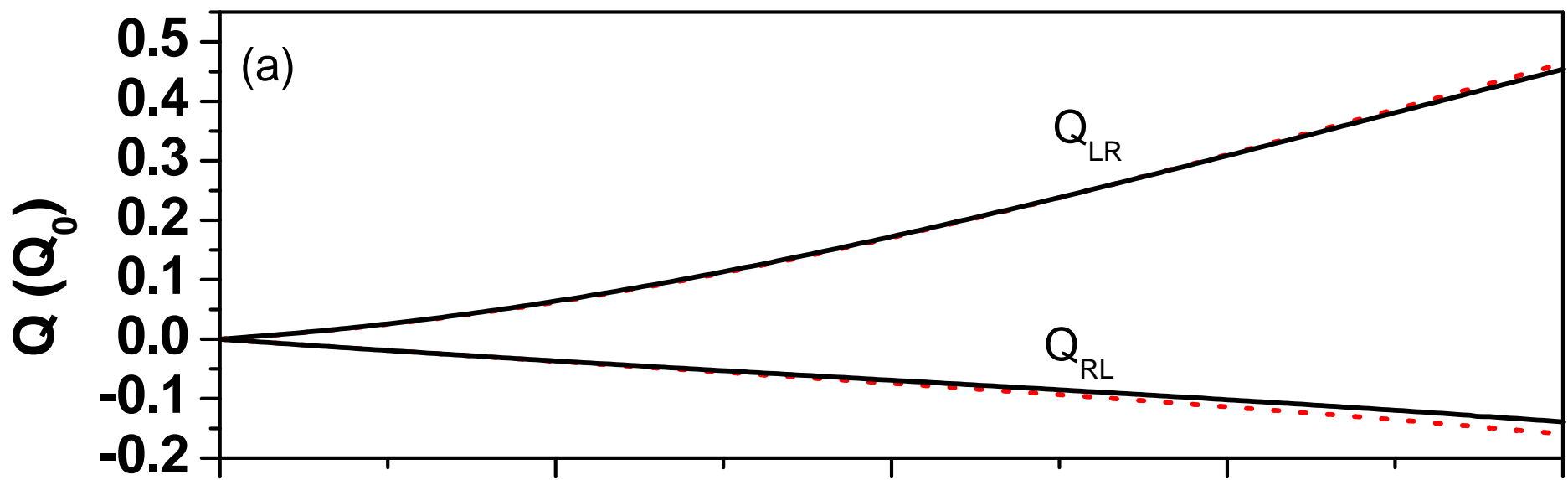
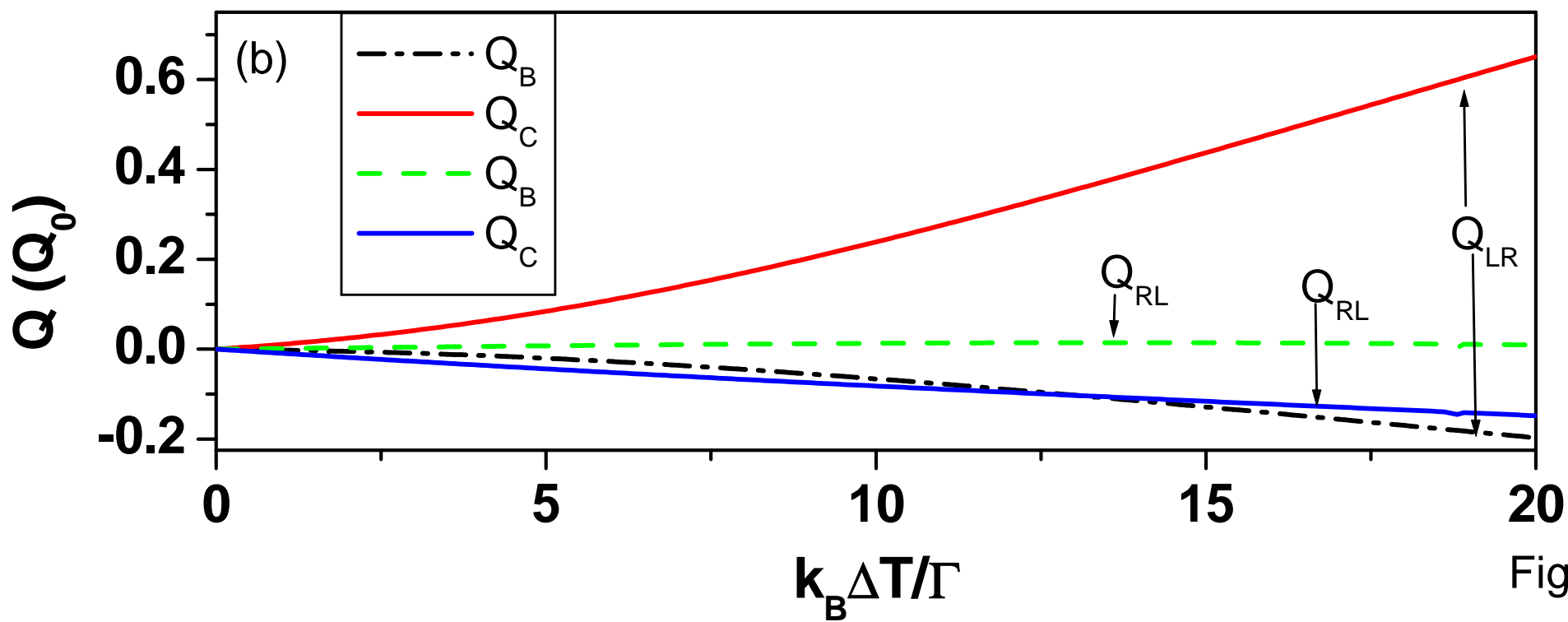
Fig. 5. Heat current as functions of $\Delta T$ for
$E_C=E_F+3\Delta E/5$. The Other parameters are the same as those
in Fig. 4.

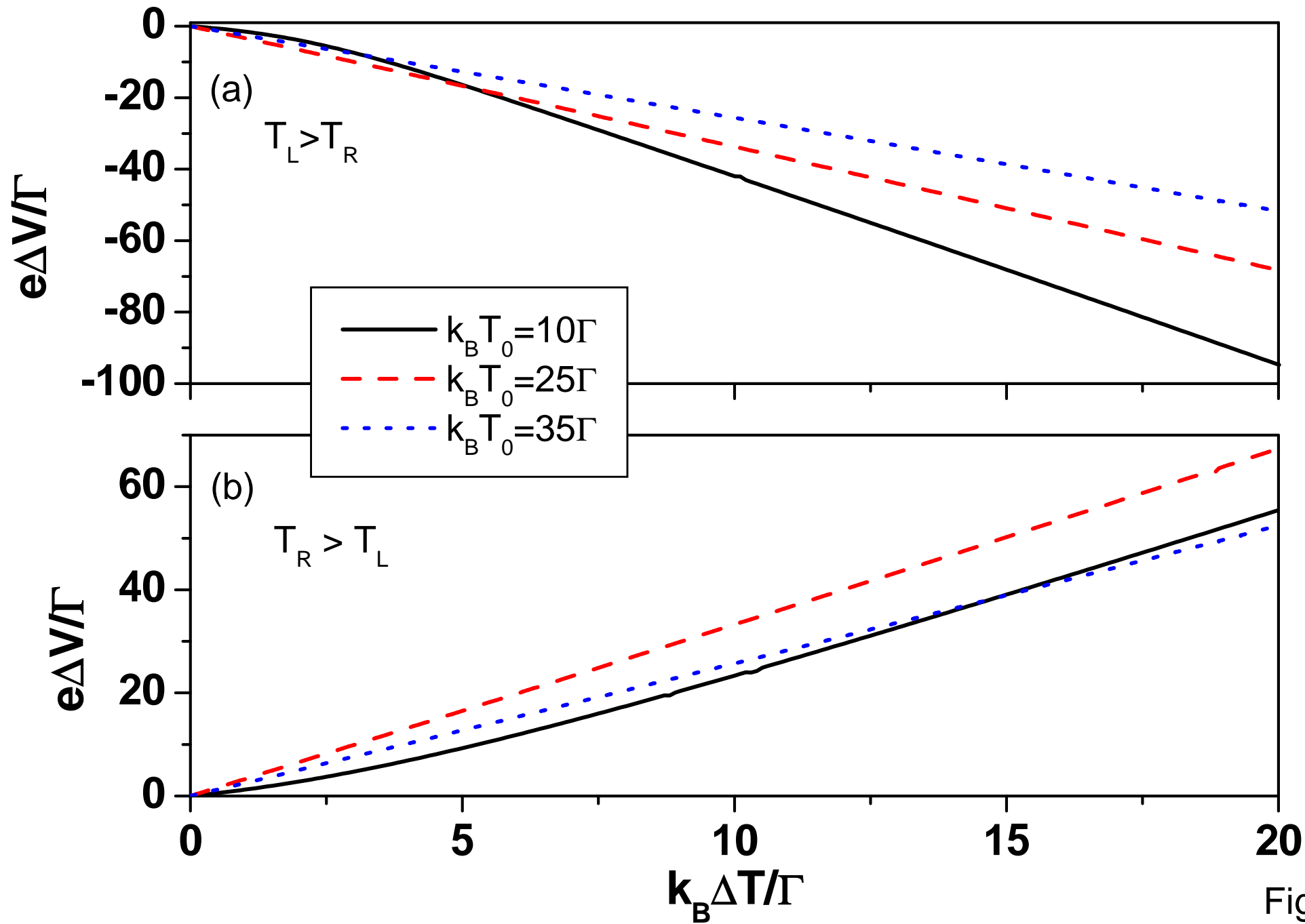
Fig. 6. Electrochemical potential as functions of $\Delta T$ for
different values of $k_BT_0$. The other parameters are the same as
those in Fig. 5.

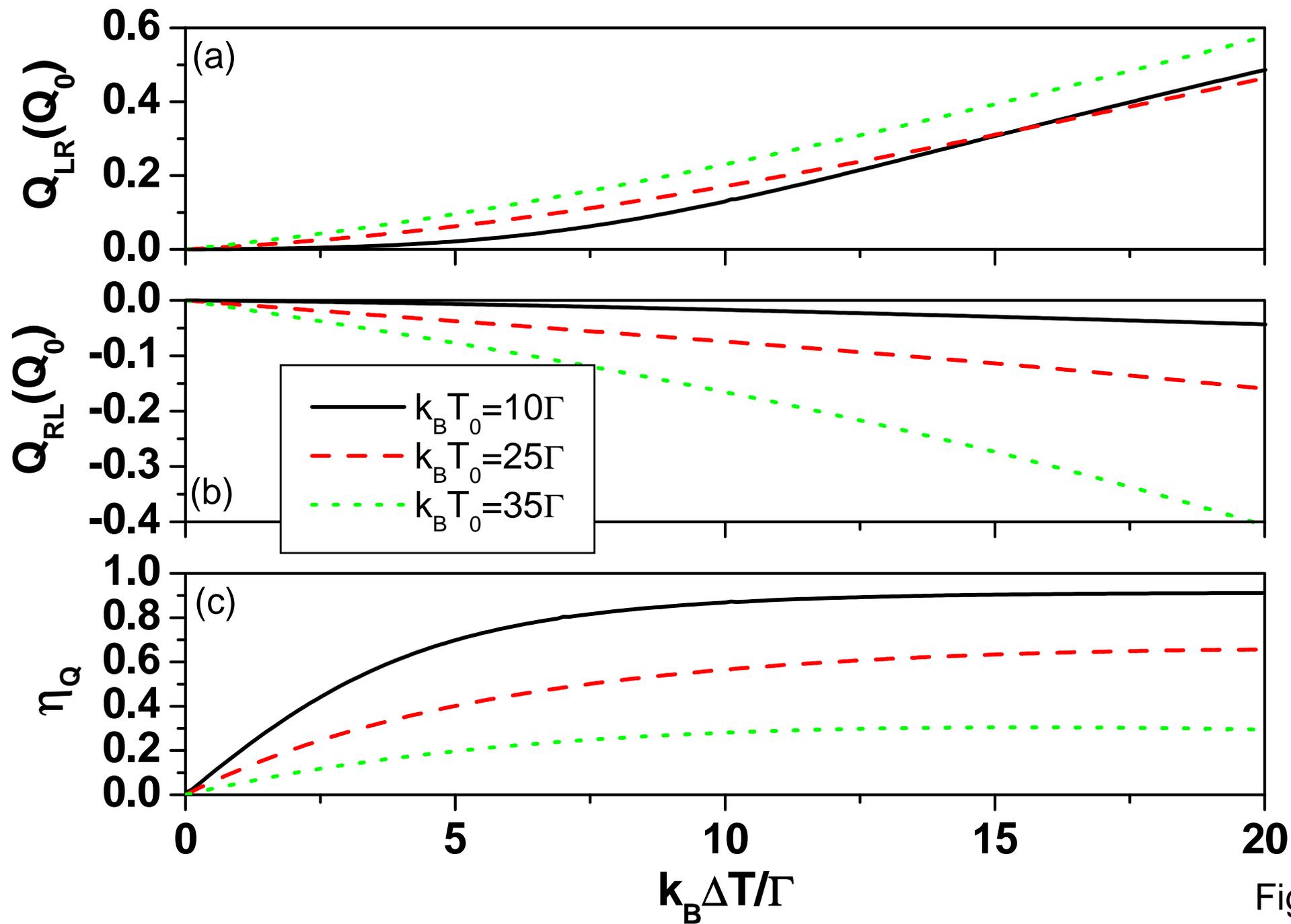
Fig. 7. Heat current and rectification efficiency as functions of
$\Delta T$ for different values of $k_BT_0$. The curves are one to
one correspondence to those of Fig. 6

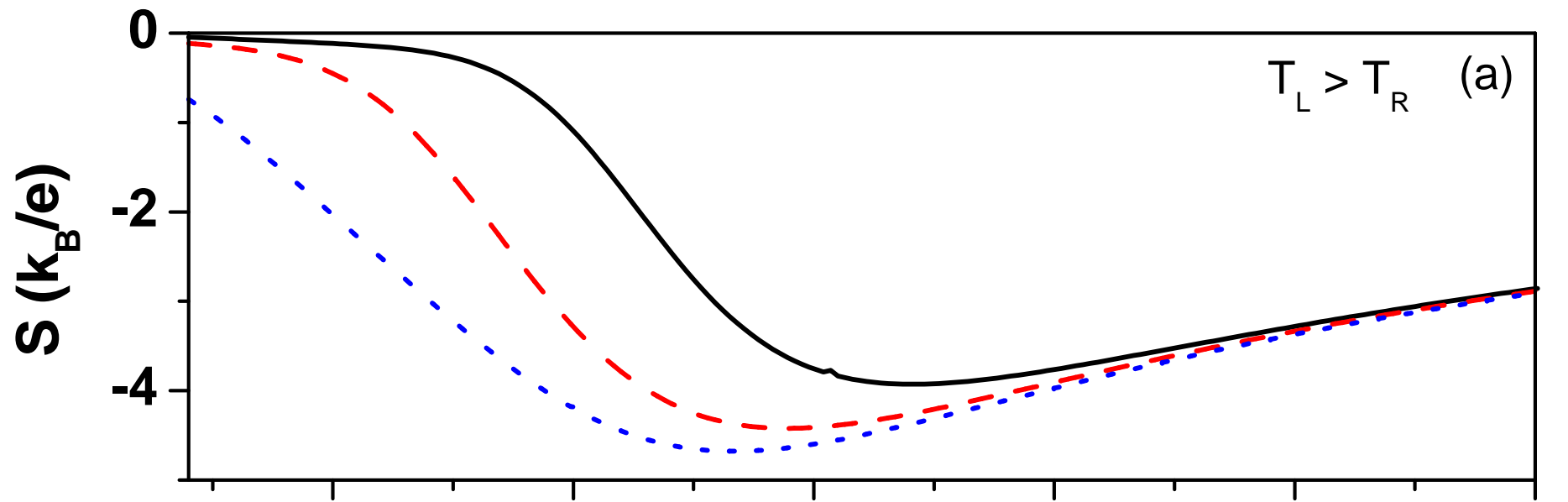
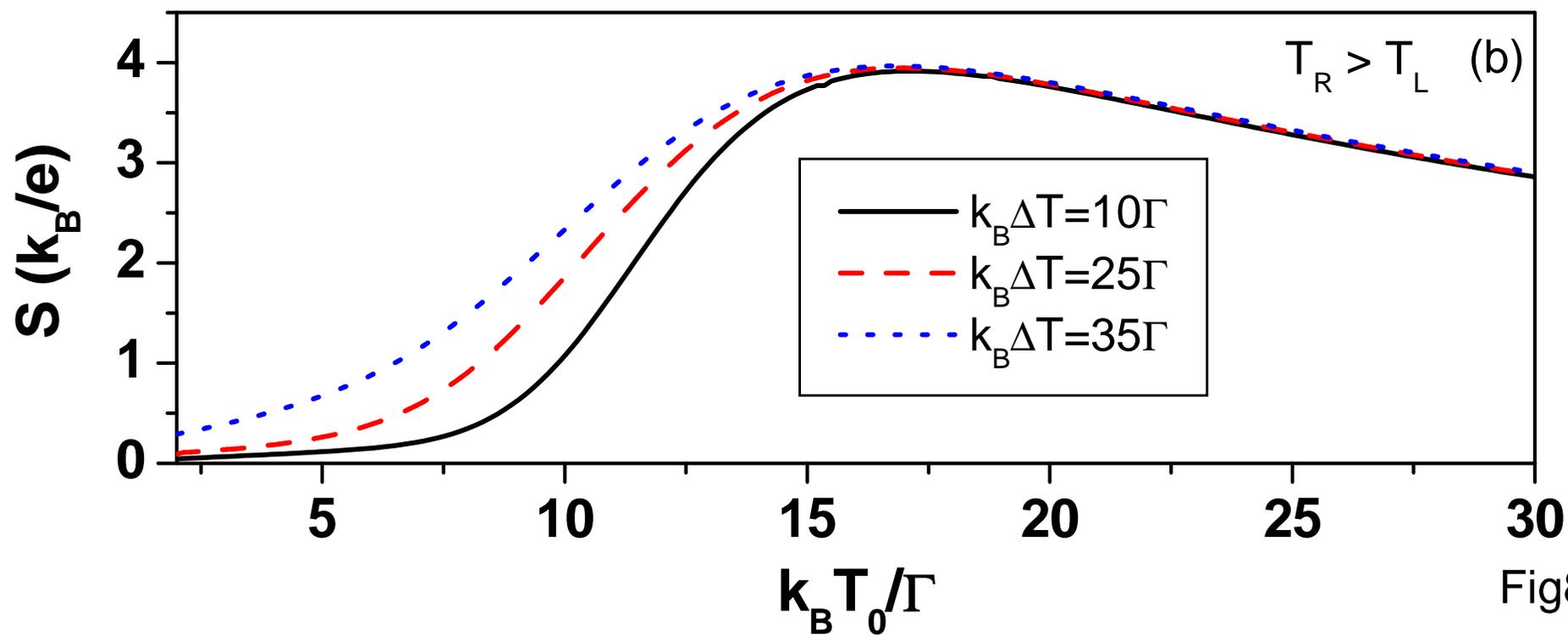
Fig. 8. Thermal power as functions of $k_BT_0 $ for the different
values of $\Delta T$. The other parameters are the same as those
in Fig. 5.


\begin{thebibliography}{50}

[1]A. J. Minnich, M. S. Dresselhaus, Z. F. Ren and G. Chen, Energy
Environ Sci, \textbf{2},  (2009) 466.\\

[2]G. Mahan, B. Sales and J. Sharp, Physics Today, \textbf{50},
(1997) 42.\\

[3]R. Venkatasubramanian, E. Siivola,T. Colpitts,B. O'Quinn,
Nature  \textbf{413}, (2001) 597.\\

[4]A. I. Boukai, Y. Bunimovich, J. Tahir-Kheli, J. K. Yu, W. A.
Goddard III and J. R. Heath, Nature, \textbf{451}, (2008) 168.\\

[5]T. C. Harman, P. J. Taylor, M. P. Walsh, B. E. LaForge, Science
\textbf{297}, (2002) 2229.\\


[6]K. F. Hsu,S. Loo,F. Guo,W. Chen,J. S. Dyck,C. Uher, T. Hogan,
E. K. Polychroniadis,M. G. Kanatzidis, Science \textbf{303},
(2004) 818.\\

[7]A. Majumdar, Science \textbf{303}, (2004) 777.\\

[8]G. Chen, M. S. Dresselhaus, G. Dresselhaus, J. P. Fleurial and
T. Caillat, International Materials Reviews, \textbf{48}, (2003)
45.\\

[9]Y. M. Lin and M. S. Dresselhaus, Phys. Rev. B \textbf{68},
 (2003) 075304.\\

[10]D. M. T. Kuo, Jpn. J. Appl. Phys. \textbf{48} (2009) 125005.\\

[11]C. Starr, J. Appl. Phys. \textbf{7}, (1936) 15.\\

[12]M. Terraneo, M. Peyrard, G. Casati, Phys. Rev. Lett.
\textbf{88},  (2002) 094302.\\

[13]Baowen Li, L. Wang and G. Casati, Phys. Rev. Lett.
\textbf{93}, (2004) 184301.\\

[14]B. Hu, L. Yang and Y. Zhang, Phys. Rev. Lett. \textbf{97},
(2006) 124302.\\

[15]G. Casati, C. Mejia-Monasterio and T. Prosen, Phys. Rev. Lett.
\textbf{98}, (2007) 104302.\\

[16]R. Scheibner, M. Konig, D. Reuter, A. D. Wieck, C. Gould, H.
Buhmann and L. W. Molenkamp,  New. J. Phys. \textbf{10}, (2008)
083016.\\

[17]X. O. Chen, B. Dong and X. L. Lei, Chin. Phys. Lett.
\textbf{25}. (2008) 3032.\\

[18]D. M. T. Kuo and Y. C. Chang, Phys. Rev. Lett. \textbf{99},
(2007) 086803.\\

[19]H. Haug and A. P. Jauho, \emph{Quantum Kinetics in Transport
and Optics of Semiconductors }(Springer, Heidelberg, 1996).\\

[20]M. Krawiec and K. I. Wysokinski, Phys. Rev. B \textbf{75},
(2007) 155330.\\

[21]B. Dong and X. L. Lei, J. Phys. Condens. Matter \textbf{14}
(2002) 11747.\\

[22]D. M. T. Kuo, Jpn. J. Appl. Phys. \textbf{47}, (2008) 8291.\\

[23]Y. C. Chang and D. M. T. Kuo, Phys. Rev. B \textbf{77}, (2008)
245412.\\

[24]D. Segal, Phys. Rev. B \textbf{73}, (2006) 205415.\\


[25]P. Murphy, S. Mukerjee and J. Moore, Phys. Rev. B \textbf{78},
(2008) 161406.\\


[26]K. H. Chen, C. Y. Chien and P. W. Li, Nanotechnology 21,
(2010) 055302.\\

[27]M. Krawiec and M. Jalochowski, Physica Status Solidi B.
\textbf{244}, (2007) 2464.\\















\end{thebibliography}
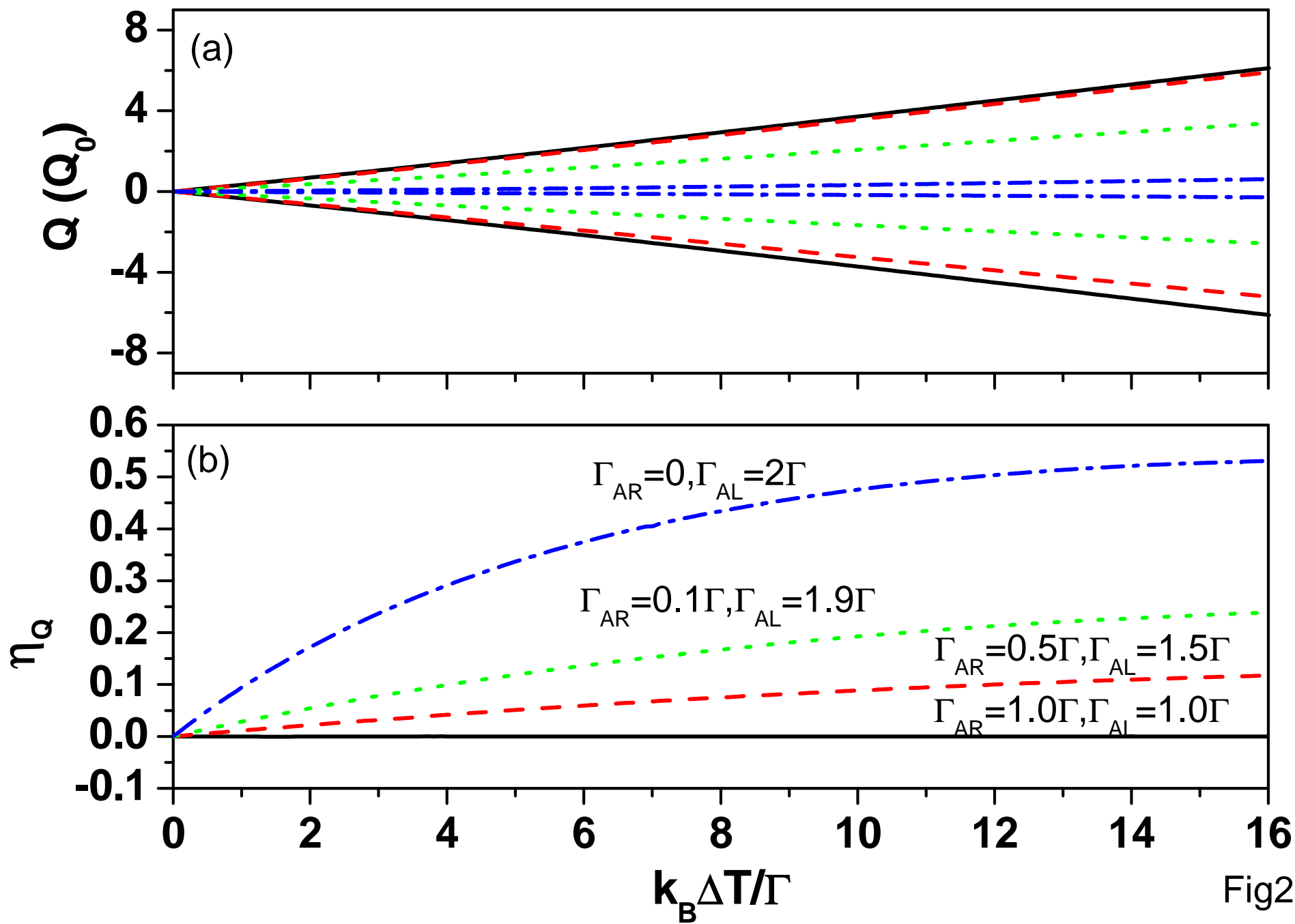
\end{document}